\documentclass[notitlepage,amsmath,amsfonts,amssymb,pra,reprint,floatfix]{revtex4-1}
\usepackage{graphicx}
\usepackage{bm}
\usepackage{color}
\begin{document}
\title{Spin-Projected Generalized Hartree-Fock as a Polynomial of Particle-Hole Excitations}
\author{Thomas M. Henderson}
\affiliation{Department of Chemistry, Rice University, Houston, TX 77005-1892}
\affiliation{Department of Physics and Astronomy, Rice University, Houston, TX 77005-1892}

\author{Gustavo E. Scuseria}
\affiliation{Department of Chemistry, Rice University, Houston, TX 77005-1892}
\affiliation{Department of Physics and Astronomy, Rice University, Houston, TX 77005-1892}
\date{\today}

\begin{abstract}
The past several years have seen renewed interest in the use of symmetry-projected Hartree-Fock for the description of strong correlations.  Unfortunately, these symmetry-projected mean-field methods do not adequately account for dynamic correlation.  Presumably, this shortcoming could be addressed if one could combine symmetry-projected Hartree-Fock with a many-body method such as coupled cluster theory, but this is by no means straightforward because the two techniques are formulated in very different ways.  However, we have recently shown that the singlet $S^2$-projected unrestricted Hartree-Fock wave function can in fact be written in a coupled cluster-like wave function: that is, the spin-projected unrestricted Hartree-Fock wave function can be written as a polynomial of a double-excitation operator acting on some closed-shell reference determinant.  Here, we extend this result and show that the spin-projected generalized Hartree-Fock wave function (which has both $S^2$ and $S_z$ projection) is likewise a polynomial of low-order excitation operators acting on a closed-shell determinant, and provide a closed-form expression for the resulting polynomial coefficients.  The spin projection of the generalized Hartree-Fock wave function introduces connected triple and quadruple excitations which are absent when spin-projecting an unrestricted Hartree-Fock determinant.  We include a few preliminary applications of the combination of this spin-projected Hartree-Fock and coupled cluster theory to the Hubbard Hamiltonian, and comment on generalizations of the methodology.  Results here are not for production level, but a similarity-transformed theory that combines the two offers the promise of being accurate for both weak and strong correlation, and may offer significant improvements in the intermediate correlation regime where neither projected Hartree-Fock nor coupled cluster is particularly accurate.
\end{abstract}
\maketitle

\section{Introduction}
Symmetry-projected Hartree-Fock \cite{Lowdin55c,Ring80,Blaizot85} has been used for many years in nuclear physics as a tool for the description of strong correlation (see, e.g., Ref. \onlinecite{Schmid2004}) and for the past several years spin-projected Hartree-Fock \cite{PHF} has seen increasing use in the computational quantum chemistry community as well.\cite{Rivero2013,Rivero2013b,JimenezHoyos2013,Stein2014,JimenezHoyos2014,RodriguezGuzman2014}  The idea is beguilingly simple.  In the presence of strong correlations, the Hartree-Fock method predicts its own failure by breaking some or all of the symmetries of the Hamiltonian.  Thus, for example, while restricted Hartree-Fock (RHF) respects both $S^2$ and $S_z$ spin symmetries, the unrestricted Hartree-Fock (UHF) wave function allows $S^2$ spin symmetry to break, the generalized Hartree-Fock (GHF) wave function allows both $S^2$ and $S_z$ spin symmetries to break, and the Bardeen-Cooper-Schrieffer (BCS) and Hartree-Fock-Bogoliubov (HFB) wave functions allow particle number symmetry to break (and, in the case of HFB, may also allow for spin symmetry breaking).  By breaking symmetry, Hartree-Fock gains variational flexibility and provides a better energy at the cost of a perhaps less physical wave function.

One could, however, retain only the component of the broken-symmetry determinant which has good symmetries, i.e. one could project the broken-symmetry Hartree-Fock determinant back onto symmetry eigenstates.  The resulting wave function is no longer a single determinant, though it can be expressed as a linear combination of non-orthogonal broken-symmetry determinants.  This projected Hartree-Fock (PHF) wave function has the convenient properties that it retains the energetic benefits of broken-symmetry Hartree-Fock but has good symmetries; it is also parameterized in terms of a single determinant, which is computationally convenient.  In practice a variation after projection approach is preferred, in which the determinant is chosen to variationally optimize the projected energy (as opposed to a projection after variation approach, in which the determinant is chosen to minimize the mean-field energy and is subsequently projected).  An important benefit of this variation after projection technique is that the underlying determinant always breaks symmetry, resulting in smooth potential energy curves.

While PHF has had some significant successes for the description of strong correlation, it is much less successful at capturing dynamic correlation.  Presumably, the fact that the wave function is based on an underlying single determinant is too severe a limitation.  Adding dynamic correlation to PHF has been a long-standing goal, and one which is beginning to come to fruition.\cite{VanVoorhis2014,Tenno2016a,Tenno2016b,WahlenStrothman2016,Hermes2017,Qiu2017}

It is worth taking a moment to understand why adding dynamic correlation to PHF is so challenging.  The basic problem is that the PHF wave function is not readily expressible as a (hopefully short) combination of single determinants in some form of active space.  On the one hand, this is a real strength of PHF: one need not define an active space at all (but can if one wishes).  On the other hand, the complicated nature of the PHF wave function makes it challenging to use PHF in combination with traditional multi-reference perturbation theory or coupled-cluster theory.

Fortunately, an alternative exists.\cite{Piecuch1996,Degroote2016,Qiu2016,Qiu2017}  One can parameterize the broken-symmetry determinant as a Thouless transformation\cite{Thouless1960} of a symmetry-adapted reference determinant; this permits the symmetry projection to be done analytically, resulting in an expression for the PHF wave function in terms of particle-hole excitations out of a reference.  Written this way, one can hope to combine PHF with more traditional wave function techniques which use a similar form of the wave function.

There is, of course, no such thing as a free lunch.  In this case, the important limitation is that different symmetries being projected yield different expressions for the PHF wave function.  Moreover, different quantum numbers of the same symmetry may do likewise.  So far as we are aware, only the particle-hole representations of the singlet spin-projected UHF (SUHF) wave function,\cite{Piecuch1996,Qiu2017} the parity projected wave function in the Lipkin Hamiltonian,\cite{WahlenStrothman2016} and the number projected HFB wave function\cite{Ring80,Dukelsky2003,Degroote2016} are known, at least among symmetries relevant for molecular systems.  We should note that the literature has focused on the number projected BCS (PBCS) case, but the number projection of HFB would be done in precisely the same way.

This manuscript adds singlet spin-projected generalized Hartree-Fock (SGHF) to the list.  By allowing all spin symmetries to break, GHF has the attractive property that it is always size consistent and size extensive\cite{JimenezHoyos2011} (though this is not true of SGHF or indeed of any PHF state, the extensive component of which is degenerate with the underlying broken-symmetry determinant).  Although GHF solutions are, on the whole, rare, SGHF is almost always different from (and variationally superior to) SUHF.  Presumably, by being variationally superior to SUHF, SGHF lends itself that little bit more readily to combination with methods which seek to add dynamic correlation.  To this end, we show a few preliminary calculations which combine SGHF with variational coupled cluster theory and compare to the combination of SUHF with variational coupled cluster.  We show results only for the Hubbard Hamiltonian,\cite{Hubbard1963} where tuning the strength of the interaction allows us to proceed smoothly from weak to strong correlation and where our previous results \cite{Qiu2017} show that the combination of SUHF with variational coupled cluster benefits enormously from the inclusion of triple excitations which, as we shall see, are built into the SGHF wave function already.  We will also discuss the combination of spin projection and number projection, to indicate how it might be done.  We emphasize, however, that the main point of this paper is to introduce the particle-hole representation of the SGHF wave function.

\section{The SGHF Wave Function}
Here, we will derive an expression for the SGHF wave function in terms of particle-hole excitations out of some RHF reference determinant.  We will follow the basic scheme explained in Ref. \onlinecite{Qiu2017}.  The derivation is a little tedious, but our main result is contained in Eqns. \ref{Eqn:FinalSGHF} - \ref{Eqn:Lambda}.

We will start by writing a representation of the GHF determinant as a particle-hole Thouless transformation out of the RHF reference:
\begin{equation}
|\mathrm{GHF}\rangle = \mathrm{e}^{T_1 + U_0 + V_+ + W_-} |\mathrm{RHF}\rangle
\end{equation}
where we have separated the Thouless transformation into different components.  Specifically, we have defined
\begin{subequations}
\begin{align}
T_1 &= \sum t_i^a \, E_{ai},
\\
U_0 &= \sum u_i^a \, S_{ai}^{0},
\\
V_+ &= \sum v_i^a \, S_{ai}^{+},
\\
W_- &= \sum w_i^a \, S_{ai}^{-},
\end{align}
\end{subequations}
where the excitation operators are
\begin{subequations}
\begin{align}
E_{ai} &= \left(c_{a_\uparrow}^\dagger \, c_{i_\uparrow} + c_{a_\downarrow}^\dagger \, c_{i_\downarrow}\right),
\\
S_{ai}^{0} &= \left(c_{a_\uparrow}^\dagger \, c_{i_\uparrow} - c_{a_\downarrow}^\dagger \, c_{i_\downarrow}\right),
\\
S_{ai}^{+} &= c_{a_\uparrow}^\dagger \, c_{i_\downarrow},
\\
S_{ai}^{-} &= c_{a_\downarrow}^\dagger \, c_{i_\uparrow},
\end{align}
\label{Def:Operators}
\end{subequations}
and where $i$ indexes spatial orbitals occupied in $|\mathrm{RHF}\rangle$ and $a$ indexes spatial orbitals unoccupied in $|\mathrm{RHF}\rangle$.  Note that $T_1$ preserves the spin quantum numbers $s$ and $s_z$, while $U_0$ turns a singlet state into a triplet state with $s_z = 0$, and $V_+$ and $W_-$ not only turn a singlet into a triplet but also increase or decrease $s_z$ by 1, respectively.  We use the notation $S_z$ to mean the spin operator and $s_z$ to denote its eigenvalues.

We should point out two limitations of the particle-hole Thouless transformation we have adopted.  First, and more seriously, we can only define it for states $|\mathrm{RHF}\rangle$ and $|\mathrm{GHF}\rangle$ which are not orthogonal, and it becomes numerically ill-conditioned if $|\mathrm{GHF}\rangle$ and $|\mathrm{RHF}\rangle$ have too little overlap.  Second, we have written the GHF determinant in a way which does not preserve its normalization.  Since the spin projection changes the normalization in any event, this is not a real limitation, but the reader should be aware that neither the GHF nor the SGHF wave functions are properly normalized in this work (though they are normalizable).

On the other hand, the particle-hole Thouless transformation has a few significant advantages.  First, the various operators all commute because they are all pure excitation operators.  Second, the coefficients of a given excitation level are easily computed.

The SGHF wave function can be extracted from the GHF determinant by finding the component which is rotationally invariant in spin space; this totally symmetric component has $s = s_z = 0$ and can be computed by \cite{PHF}
\begin{subequations}
\begin{align}
|\mathrm{SGHF}\rangle 
 &= \int_0^{2\pi} \frac{\mathrm{d}\gamma}{2\pi} \int_0^\pi \frac{\mathrm{d}\beta \, \sin(\beta)}{2} \int_0^{2\pi} \frac{\mathrm{d}\alpha}{2\pi} \, R(\Omega)  |\mathrm{GHF}\rangle
\\
 &= \frac{1}{2} \, P_{s_z = 0} \, \int_0^\pi \mathrm{d}\beta \, \sin(\beta)  \, \mathrm{e}^{\mathrm{i} \, \beta \, S_y} \, P_{s_z = 0} |\mathrm{GHF}\rangle,
\nonumber
\\
R(\Omega) &=  \mathrm{e}^{\mathrm{i} \, \gamma \, S_z} \, \mathrm{e}^{\mathrm{i} \, \beta \, S_y} \, \mathrm{e}^{\mathrm{i} \, \alpha \, S_Z}.
\end{align}
\end{subequations}
Here, $P_{s_z = 0}$ is the projector onto states with $s_z = 0$.  Because the $T_1$ component of the Thouless transformation carrying RHF to GHF commutes with the spin rotation operator $R(\Omega)$, it is a spectator in the various projections and integrations of the foregoing equation, and we will omit it for simplicity.

Before we can go further, we need the portion of the GHF wave function which has $s_z = 0$.  This is readily obtained.  Noting that the projection operator commutes with $U_0$, we can write
\begin{subequations}
\begin{align}
P_{s_z = 0} |\mathrm{GHF}\rangle
 &= \mathrm{e}^{U_0} \, P_{s_z = 0} \, \mathrm{e}^{V_+ + W_-} |\mathrm{RHF}\rangle
\\
 &= \mathrm{e}^{U_0} \, P_{s_z = 0} \, \sum_{m,n} \frac{1}{m!} \, \frac{1}{n!} V_+^m \, W_-^n |\mathrm{RHF}\rangle
\\
 &= \mathrm{e}^{U_0} \, \sum_n \, \frac{1}{(n!)^2} \, \left(V_+ \, W_-\right)^n |\mathrm{RHF}\rangle,
\end{align}
\end{subequations}
where we have simply used the fact that the projection onto $s_z = 0$ requires us to increase the $z$-component of spin using $V_+$ precisely as many times as we decrease it using $W_-$.  This is the same logic as used to derive the number projection in number-projected BCS.\cite{Degroote2016}

Now we need to act the rotation operator $\exp(\mathrm{i} \, \beta \, S_y)$.  It is convenient to use
\begin{subequations}
\begin{align}
\mathrm{e}^{\mathrm{i} \, \beta \, S_y} \, \mathcal{O} \, |\mathrm{RHF}\rangle 
 &= \mathrm{e}^{\mathrm{i} \, \beta \, S_y} \, \mathcal{O} \, \mathrm{e}^{-\mathrm{i} \, \beta \, S_y} \, \mathrm{e}^{\mathrm{i} \, \beta \, S_y} \, |\mathrm{RHF}\rangle 
\\
 &= \tilde{\mathcal{O}} |\mathrm{RHF}\rangle,
\end{align}
\end{subequations}
where
\begin{equation}
\tilde{\mathcal{O}} = \mathrm{e}^{\mathrm{i} \, \beta \, S_y} \, \mathcal{O} \, \mathrm{e}^{-\mathrm{i} \, \beta \, S_y}.
\end{equation}
We have used the fact that the RHF wave function is an eigenfunction of $S_y$ with eigenvalue 0.  We thus need transformed versions of $U_0$, $V_+$, and $W_-$, which we will respectively call $\tilde{U}$, $\tilde{V}$, and $\tilde{W}$.  Writing $S_y$ in terms of the fermionic operators $c$ and $c^\dagger$ and using the commutator expansion yields
\begin{subequations}
\begin{align}
\tilde{U}
 &= \cos(\beta) \, U_0 - \sin(\beta) \, \left(U_+ + U_-\right),
\\
\tilde{V}
 &= \frac{1}{2} \, \sin(\beta) \, V_0 + \frac{1 + \cos(\beta)}{2} \, V_+ - \frac{1 - \cos(\beta)}{2} \, V_-,
\\
\tilde{W}
 &= \frac{1}{2} \, \sin(\beta) \, W_0 + \frac{1 + \cos(\beta)}{2} \, W_- - \frac{1 - \cos(\beta)}{2} \, W_+.
\end{align}
\end{subequations}
We have defined additional operators in analogy with $U_0$, $V_+$, and $W_-$.  Generically, the operators $\Theta_k$ are given by
\begin{equation}
\Theta_k = \sum \theta_i^a \, S_{ai}^k
\end{equation}
where $k$ is $0$, $+$, or $-$ and where $\theta$ = $u$, $v$, or $w$ when $\Theta$ is $U$, $V$, or $W$, respectively.

At this point, the SGHF wave function could be written as
\begin{equation}
|\mathrm{SGHF}\rangle = \sum_N C_N |\mathrm{RHF}\rangle
\end{equation}
where $N$ indexes the total excitation level.  The $N$-fold excitation operator $C_N$ is given by
\begin{align}
C_N &= \frac{1}{2} \, \sum_{M=0}^{N/2} \, \frac{1}{(N-2M)! \, M! \, M!} 
\\
  &\times P_{s_z=0} \, \int_0^\pi \mathrm{d}\beta \, \sin(\beta) \, \tilde{U}^{N-2M} \, \tilde{V}^M \, \tilde{W}^M
\nonumber
\end{align}
where the action of the projection operator is understood to mean taking the $\Delta s_z = 0$ part of the operator string.

One can introduce the definitions of $\tilde{U}$, $\tilde{V}$, and $\tilde{W}$ and use multinomial coefficients to expand the expressions for the rotated operators:
\begin{widetext}
\begin{subequations}
\begin{align}
\frac{\tilde{U}^{N-2M}}{(N-2M)!} &= \sum_{i=0}^{N-2M} \, \sum_{j=0}^{N-2M-i} \frac{U_0^{N-2M-i-j}}{(N-2M-i-j)!} \, \frac{U_+^i}{i!} \, \frac{U_-^j}{j!} \, \cos(\beta)^{N-2M-i-j} \, \left[-\sin(\beta)\right]^{i+j},
\\
\frac{\tilde{V}^M}{M!} &= \sum_{k=0}^M \, \sum_{l=0}^{M-k} \frac{V_0^{M-k-l}}{(M-k-l)!} \, \frac{V_+^k}{k!} \, \frac{V_-^l}{l!} \, \left[\frac{1}{2} \, \sin(\beta)\right]^{M-k-l} \, \left[\frac{1+\cos(\beta)}{2}\right]^k \, \left[-\frac{1-\cos(\beta)}{2}\right]^l,
\\
\frac{\tilde{W}^M}{M!} &= \sum_{m=0}^M \, \sum_{n=0}^{M-n} \frac{W_0^{M-m-n}}{(M-m-n)!} \, \frac{W_+^m}{m!} \, \frac{W_-^n}{n!} \, \left[\frac{1}{2} \, \sin(\beta)\right]^{M-m-n} \, \left[\frac{1+\cos(\beta)}{2}\right]^n \, \left[-\frac{1-\cos(\beta)}{2}\right]^m.
\end{align}
\end{subequations}
This means that
\begin{align}
C_N &= \sum_{M=0}^{N/2} \, \sum_{i=0}^{N - 2 M} \, \sum_{j = 0}^{N - 2 M - i} \, \sum_{k=0}^M \, \sum_{l=0}^{M-k} \, \sum_{m=0}^M \, \sum_{n=0}^{M-m} \frac{\left(-1\right)^{k+n}}{2^{2M+1}} \, \mathcal{I}(N - 2M - i - j,M + i + k - l,M + i + m - n)
\label{Eqn:DefCN}
\\
 &\qquad\qquad \times \delta_{i+k+m,j+l+n} \, \frac{U_0^{N-2M-i-j}}{(N-2M-i-j)!} \, \frac{U_+^i}{i!} \, \frac{U_-^j}{j!} \, \frac{V_0^{M-k-l}}{(M-k-l)!} \, \frac{V_+^k}{k!} \, \frac{V_-^l}{l!} \, \frac{W_0^{M-m-n}}{(M-m-n)!} \, \frac{W_+^m}{m!} \, \frac{W_-^n}{n!},
\nonumber
\end{align}
\end{widetext}
where the Kronecker delta arises from the $s_z = 0$ projection operator and where the integral $\mathcal{I}(p,q,r)$ is given by
\begin{equation}
\mathcal{I}(p,q,r) = \int_{-1}^1 \mathrm{d}x \, x^p \, (1+x)^q \, (1-x)^r.
\end{equation}
Here, we have written $x = \cos(\beta)$ and simplified the exponent on $\sin(\beta)$; recalling that one factor of $\sin(\beta)$ is absorbed by the change of variables, for the rest we have
\begin{equation}
\sin(\beta)^{2M+i+j-k-l-m-n} = \sin(\beta)^{2(M+i-l-n)}
\end{equation}
where we have used the fact that the Kronecker delta enforces that $k+m = j + l + n - i$.  Because the exponent on $\sin(\beta)$ is even, we have written
\begin{align}
\left[\sin(\beta)^2\right]^{M+i-l-n} &= \left[1 + \cos(\beta)\right]^{M+i-l-n}
\\
 &\times \left[1 - \cos(\beta)\right]^{M+i-l-n}.
\nonumber
\end{align}

At this point we resorted to computer algebra \cite{Mathematica}, and obtained a rather surprising result.  Although we have been unable to rigorously prove it, we find that
\begin{equation}
C_N = \sum_{2i+3j+4k=N} \lambda_{ijk} \, C_2^i \, C_3^j \, K_4^k;
\label{Eqn:DefCN2}
\end{equation}
hence
\begin{equation}
|\mathrm{SGHF}\rangle = \sum_{ijk} \lambda_{ijk} \, C_2^i \, C_3^j \, K_4^k |\mathrm{RHF}\rangle
\label{Eqn:FinalSGHF}
\end{equation}
where
\begin{subequations}
\begin{align}
C_2 &= \kappa(U,U) + \kappa(V,W),
\\
C_3 &= \frac{1}{6} \, \big[U_0 \, \left(V_+ \, W_- - V_- \, W_+\right)
\\
  &\quad  + V_0 \, \left(W_+ \, U_- - W_- \, U_+\right)
\nonumber
\\
 &\quad + W_0 \, \left(U_+ \, V_- - U_- \, V_+\right)\big],
\nonumber
\\
K_4 &= C_4 - \frac{3}{10} \, C_2^2
\\
  &= \frac{3}{5} \, \left[\kappa(U,U) \, \kappa(V,W) - \kappa(U,V) \, \kappa(U,W)\right]
\nonumber
\\
 &+ \frac{3}{20} \, \left[\kappa(V,W)^2 - \kappa(V,V) \, \kappa(W,W)\right],
\nonumber
\\
\kappa(X,Y) &= \frac{1}{6} \, X_0 \, Y_0 + \frac{1}{3} \, X_+ \, Y_- + \frac{1}{3} \, X_- \, Y_+,
\end{align}
\label{Eqn:DefC2C3K4}
\end{subequations}
and
\begin{equation}
\lambda_{ijk} = \frac{6^i}{i!} \, \frac{12^j}{j!} \, \frac{60^k}{k!} \, \frac{(i + 2j + 2k)!}{(k+j)! \, (2i + 3j + 4k + 1)!}.
\label{Eqn:Lambda}
\end{equation}
Equations \ref{Eqn:DefCN2} - \ref{Eqn:Lambda} constitute the main result of this paper.  We have checked that they yield precisely the same coefficients $C_N$ as are computed by the explicit form of Eqn. \ref{Eqn:DefCN} through 30-fold excitations (i.e. $N = 30$), though we cannot guarantee that they are correct for still higher excitation levels.  It is important to note that where SUHF contains only single and double excitations and powers of them\cite{Piecuch1996,Qiu2016,Qiu2017}, SGHF additionally has triple and quadruple excitations which arise from the additional $S_z$ projections and which may have significant energetic consequences (see below).

Table \ref{Tab:Lambda} provides numerical values for the coefficients $\lambda_{ijk}$ through octuple excitations, to give an idea of how rapidly the coefficients decay, and compares to the analogous coefficients in coupled cluster theory.  In coupled cluster, the wave function is a simple exponential:
\begin{equation}
|\mathrm{CC}\rangle = \mathrm{e}^T | \mathrm{RHF}\rangle
\end{equation}
where different levels of theory correspond to different choices of the operator $T$.  Since SGHF includes up through quadruple excitations, the natural comparison is coupled cluster with singles, doubles, triples, and quadruples (CCSDTQ).  Omitting the singles for brevity, we would have
\begin{equation}
|\mathrm{CCDTQ}\rangle = \sum \frac{1}{i!} \, \frac{1}{j!} \, \frac{1}{k!} \, T_2^i \, T_3^j \, T_4^k
\end{equation}
where $T_2$, $T_3$, and $T_4$ are double, triple, and quadruple excitation operators.  The coefficients in the SGHF wave function clearly decay much more rapidly with increasing excitation level than do the corresponding coefficients in coupled cluster.  The same is true for spin projected UHF and for number projected BCS, which have respectively
\begin{subequations}
\begin{align}
|\mathrm{SUHF}\rangle &= \sum \frac{6^i}{(2i+1)!} \, T_2^i |\mathrm{RHF}\rangle,
\\
|\mathrm{PBCS}\rangle &= \sum \frac{1}{i! \, i!} \, T_2^i |\mathrm{RHF}\rangle,
\end{align}
\end{subequations}
where both contain only double excitation operators which, like the SGHF excitation operators, are parameterized in terms of the Thouless transformation taking RHF to the broken-symmetry determinant.  This observation for number and $S^2$ spin projection forms the basis for our work on what we call attenuated coupled cluster,\cite{Gomez2017} in which we add the SUHF or PBCS polynomial to the coupled cluster exponential, thereby substantially mitigating the failures of traditional coupled cluster theory in the stongly correlated limit.  Both the SUHF and the PBCS polynomials are examples of (univariate) hypergeometric functions, as indeed is the exponential of coupled cluster theory; the SGHF polynomial is a multivariate generalization.

\begin{table}[t]
\caption{Coefficients $\lambda$ for the SGHF wave function and for the analogous coupled cluster wave function.  The excitation level is $N = 2i + 3j + 4k$.  The coefficients in coupled cluster theory are simply $1/(i! \, j! \, k!)$, and we call them $\lambda_{ijk}^\mathrm{CC}$.
\label{Tab:Lambda}}
\begin{tabular}{ccccll}
\hline\hline
$N$  & $i$  &  $j$  &  $k$ &  $\lambda_{ijk}$  & $\lambda_{ijk}^\mathrm{CC}$
\\
\hline
2 & 1 & 0 & 0 &  1 & 1
\\
\hline
3 & 0 & 1 & 0 &  1 & 1
\\
\hline
4 & 0 & 0 & 1 & 1  & 1
\\
  & 2 & 0 & 0 & 3/10 & 1/2
\\  
\hline
5 & 1 & 1 & 0 & 3/5 & 1
\\
\hline
6 & 1 & 0 & 1 & 3/7 & 1
\\
  & 0 & 2 & 0 & 6/35 & 1/2
\\
  & 3 & 0 & 0 & 3/70 & 1/6
\\
\hline
7 & 0 & 1 & 1 & 3/14 & 1
\\
  & 2 & 1 & 0 & 9/70 & 1/2
\\
\hline
8 & 0 & 0 & 2 & 5/84 & 1/2
\\
  & 2 & 0 & 1 & 1/14 & 1/2
\\
  & 1 & 2 & 0 & 1/14 & 1/2
\\
  & 4 & 0 & 0 & 1/280 & 1/24
\\
\hline\hline
\end{tabular}
\end{table}

It is not terribly obvious that $C_2$, $C_3$, and $K_4$ really are singlet operators (that is, operators expressible in terms of the unitary group generators $E_{ai}$ of Eqn. \ref{Def:Operators}).  The form of $\kappa(X,Y)$ is the same as appears in the SUHF polynomial, however, and with a little algebra one finds that
\begin{align}
\kappa(X,Y) = -\frac{1}{12} \, \sum_{ijab} \big(&x_i^a \, y_j^b + 2 \, x_i^b \, y_j^a
\\
 +& y_i^a \, x_j^b + 2 \, y_i^b \, x_j^a\big) \, E_{ai} \, E_{bj}
\nonumber
\end{align}
where $i$ and $j$ index occupied spatial orbitals and $a$ and $b$ index virtual spatial orbitals.  Since $\kappa(X,Y)$ is obviously a singlet, it follows that both $C_2$ and $K_4$ are indeed singlet operators.  We do not at present have a clean demonstration that $C_3$ actually is a singlet operator, though of course it must be.

A few brief comments are in order.  First, we use $K_4$ instead of $C_4$ in Eqn. \ref{Eqn:FinalSGHF} simply because in SUHF, $C_3 = 0$ and $C_4 = 3/10 \, C_2^2$; thus, $K_4$ is the new quadruply-excited part of the SGHF wave function.  Second, both $C_3$ and $K_4$ are important parts of the SGHF wave function; if they are neglected, one essentially falls back to SUHF.  Finally, if one sets $\bm{v} = \bm{w} = \bm{0}$, so that the Thouless transformation forms only a UHF wave function, our result properly reduces to SUHF.

To validate our SGHF expression, we have implemented SGHF in a full configuration interaction code, using the conjugate gradients method to optimize the parameters $t_i^a$, $u_i^a$, $v_i^a$, and $w_i^a$.  Note that $\bm{u} = \bm{v} = \bm{w} = \bm{0}$ is a solution of the SGHF equations which simply corresponds to RHF, and $\bm{v} = \bm{w} = \bm{0}$ is also a solution and corresponds to SUHF.  We must therefore provide nonzero initial guesses for $\bm{u}$, $\bm{v}$, and $\bm{w}$.  Our results are unfortunately somewhat sensitive to these initial guesses, and this is complicated further by the difficulty in finding GHF solutions for use as an initial guess, since many systems do not spontaneously produce GHF determinants.  We have found many GHF solutions in the periodic Hubbard Hamiltonian, but most are orthogonal to the RHF ground state and hence cannot be reached by a non-unitary Thouless transformation of the sort we haved used in our derivation.

Nevertheless, we have tested our polynomial for the 6-site Hubbard model at half-filling by carrying out SGHF calculations with real molecular orbitals (hence real $\bm{t}$, $\bm{u}$, $\bm{v}$ and $\bm{w}$) using the standard technique of integrating over the symmetry coherent states\cite{PHF} and compared to results using our SGHF polynomial.  The energies agree to the convergence tolerance of the respective calculations (in this case, to better than six decimal places).  Though the half-filled six-site Hubbard model does not test polynomial coefficients for excitations higher than $C_6$, it serves to validate both the coefficients to that order and the procedure we used to derive these coefficients.  Note that our polynomial does not assume real wave function parameters and is equally valid for complex Thouless transformations; our code, however, is limited to a real Thouless transformation acting on a real RHF wave function.  This means that the underlying broken-symmetry GHF determinant is limited to coplanar spins.  Similarly, while we have tested it for the Hubbard Hamiltonian, our derivation is universal and applies to \textit{any} generalized Hartree-Fock wave function, provided only that the broken-symmetry GHF determinant is not orthogonal to the RHF determinant used as a reference for the Thouless transformation.

\subsection{Discussion}
It may be illuminating to explain how we obtained the formula for $\lambda_{ijk}$ given in Eqn. \ref{Eqn:Lambda}.

First, then, we obtained the values of the coefficients by finding the explicit form of $C_N$ using Eqn. \ref{Eqn:DefCN} and assuming it had the factorized form of Eqn. \ref{Eqn:DefCN2}.  We compared unique operator strings on each side of the foregoing equation (e.g. $U_0^4 \, V_+ \, W_-$) and solved for the coefficients $\lambda_{ijk}$ needed to satisfy the equality.  Although we have not been able to automate this  process, we have had little difficulty in carrying it out.

Inspired by the form of the SUHF wave function, we grouped all powers of $C_2$ for a given combination $C_3^j \, K_4^k$ and observed that
\begin{widetext}
\begin{align}
|\mathrm{SGHF}\rangle
 &= \Big\{        \sum b_{0,k} \, \bar{i}_{2k}(C_2) \, K_4^k
\label{Eqn:BigSGHF}
  + C_3 \, \sum b_{1,k} \, \bar{i}_{2k+1}(C_2) \, K_4^k
\\
 &+ C_3^2 \, \sum b_{2,k} \, \big[\bar{i}_{2k+2}(C_2) + \bar{i}_{2k+3}(C_2)\big] \, K_4^k
  + C_3^3 \, \sum b_{3,k} \, \big[\bar{i}_{2k+3}(C_2) + 3 \, \bar{i}_{2k+4}(C_2)\big] \, K_4^k
\nonumber
\\
 &+ C_3^4 \, \sum b_{4,k} \, \big[\bar{i}_{2k+4}(C_2) + 6 \, \bar{i}_{2k+5}(C_2) + 3 \, \bar{i}_{2k+6}(C_2)\big] \, K_4^k
\nonumber
\\
 &+ \ldots \Big\} |\mathrm{RHF}\rangle
\nonumber
\end{align}
\end{widetext}
accurately predicted all $\lambda_{ijk}$ that we had found.  The special functions $\bar{i}_n$ are related to modified spherical Bessel functions of the first kind:
\begin{subequations}
\begin{align}
\bar{i}_n(x)
 &= \frac{i_n(\sqrt{6 x})}{(6 x)^{n/2}},
\\
 &= \frac{1}{3} \, \frac{\mathrm{d}\hfill}{\mathrm{d}x} \, \bar{i}_{n-1}(x),
\label{Eqn:Recurrence}
\\
 &= 2^n \, \sum_{m=0}^\infty \frac{(m+n)!}{m!} \, \frac{(6 \, x)^m}{(2m+2n+1)!}.
\label{Eqn:TaylorSeriesIBar} 
\end{align}
\end{subequations}
The coefficients $b_{j,k}$ turned out to be
\begin{equation}
b_{j,k} = \frac{3^j}{j!} \, \frac{15^k}{k!} \, \frac{1}{(k+j)!},
\end{equation}
though this was not initially obvious to us.

What was clear was that for a given power of $C_3$, increasing the power of $K_4$ by one is always accompanied by increasing the index or indices on the $\bar{i}_n$ by two.  By the recursion relation of Eqn. \ref{Eqn:Recurrence}, this is tantamount to taking two derivatives, which permits us to rewrite the SGHF wave function as
\begin{widetext}
\begin{align}
|\mathrm{SGHF}\rangle
 &= \Big\{\left[1 + \mathcal{X} + \frac{1}{4} \, \mathcal{X}^2 + \frac{1}{36} \, \mathcal{X}^3 + \frac{1}{576} \, \mathcal{X}^4 + \frac{1}{14400} \, \mathcal{X}^5 + \ldots\right] \, \bar{i}_0(C_2)
\label{Eqn:SGHFXForm}
\\
 &+ 3 \, \left[1 + \frac{1}{2} \, \mathcal{X} + \frac{1}{12} \, \mathcal{X}^2 + \frac{1}{144} \, \mathcal{X}^3 + \frac{1}{2880} \, \mathcal{X}^4 + \ldots\right] \, C_3 \, \bar{i}_1(C_2)
\nonumber
\\
 &+ \frac{9}{4} \, \left[1 + \frac{1}{3} \, \mathcal{X} + \frac{1}{24} \, \mathcal{X}^2 + \frac{1}{360} \, \mathcal{X}^3 + \ldots\right] \, C_3^2 \, \left[\bar{i}_2(C_2) + \bar{i}_3(C_2)\right]
\nonumber
\\
 &+ \frac{3}{4} \, \left[1 + \frac{1}{4} \, \mathcal{X} + \frac{1}{40} \, \mathcal{X}^2 + \ldots\right] \, C_3^3 \, \left[\bar{i}_3(C_2) + 3 \, \bar{i}_4(C_2)\right]
\nonumber
\\
 &+ \frac{9}{64} \, \left[1 + \frac{1}{5} \, \mathcal{X} + \frac{1}{60} \, \mathcal{X}^2 + \ldots\right] \, C_3^4 \, \left[\bar{i}_4(C_2) + 6 \, \bar{i}_5(C_2) + 3 \, \bar{i}_6(C_2)\right] + \ldots \Big\} |\mathrm{RHF}\rangle
\nonumber 
\end{align}
\end{widetext}
where we have introduced
\begin{equation}
\mathcal{X} = \frac{5}{3} \, K_4 \, \mathcal{D}_{C_2}^2
\label{Eqn:DefX}
\end{equation}
and where
\begin{equation}
\mathcal{D}_{C_2} = \frac{\mathrm{d}\hfill}{\mathrm{d}C_2}.
\end{equation}

While we had not recognized the pattern in the $b_{j,k}$, to our gratified surprise we recognized the coefficients in the power series in $\mathcal{X}$ for a given power of $C_3$.  Defining a second set of special functions $\bar{I}_n$ related to the modified (but not spherical) Bessel functions and satisfying
\begin{subequations}
\begin{align}
\bar{I}_n(x) &= \frac{I_n(2 \sqrt{x})}{x^{n/2}},
\\
 &= \frac{\mathrm{d}\hfill}{\mathrm{d}x} \bar{I}_{n-1}(x),
\label{Eqn:RecursionBigIBar}
\\
 &= \sum_{k=0}^\infty \frac{x^k}{k! \, (n+k)!},
\end{align}
\end{subequations}
we saw that we were apparently able to write the SGHF wave function in the compact and rather suggestive form
\begin{align}
|\mathrm{SGHF}\rangle 
 &= \sum \frac{1}{j!} \, C_3^j \, \bar{I}_j(\mathcal{X}) \, \mathcal{D}_{C_2}^j \, P_j\left(\frac{1}{3}\mathcal{D}_{C_2}\right)
\\
 &\qquad\qquad \times \bar{i}_0(C_2) |\mathrm{RHF}\rangle
\nonumber
\end{align}
where we have checked the sum on $j$ through $j = 4$ and where the polynomials $P_j$ are
\begin{subequations}
\begin{align}
P_0(x) &= 1,
\\
P_1(x) &= 1,
\\
P_2(x) &= 1 + x
\\
P_3(x) &= 1 + 3 \, x
\\
P_4(x) &= 1 + 6 \, x + 3 \, x^2.
\end{align}
\end{subequations}

Assuming the foregoing result to be true for all $ik$, we derived explicit expressions for the coefficients $\lambda_{ijk}$ for all $ik$ for a given $j$.  Having done so for several values of $j$, the final result of Eqn. \ref {Eqn:Lambda} was obvious.

Let us, then, be clear about what we do and do not mean to say.  We do not claim to have proven that the SGHF wave function factorizes to all excitation levels, but it does factorize at all excitation levels we have been able to check.  Likewise, though the coefficients of Eqn. \ref{Eqn:Lambda} are correct in every case we have been able to verify, they were obtained by recognizing patterns in the coefficients for low excitation levels and assuming those  patterns hold \textit{ad infinitum}.  This they need not do, and though we think it unlikely, a counterexample to either Eqn. \ref{Eqn:FinalSGHF} or Eqn. \ref{Eqn:Lambda} might eventually be found.

\section{SGHF in Terms of Tensor Invariants}
Writing the SGHF Thouless transformation in terms of the operators $U_0$, $V_+$, and $W_-$ simplifies the derivation, but comes at a cost: the three operators are treated differently in the final expression.  A more symmetric expression can be derived by considering an alternative parameterization.

Let us first write down an alternative set of triplet single excitation operators to replace $U$, $V$, and $W$:
\begin{subequations}
\begin{align}
X &= \sum_{ia} \sum_{\zeta\eta} x_i^a \, c_{a_\zeta}^\dagger \, c_{i_\eta} \, \sigma^x_{\zeta\eta},
\\
Y &= \sum_{ia} \sum_{\zeta\eta} y_i^a \, c_{a_\zeta}^\dagger \, c_{i_\eta} \, \sigma^y_{\zeta\eta},
\\
Z &= \sum_{ia} \sum_{\zeta\eta} z_i^a \, c_{a_\zeta}^\dagger \, c_{i_\eta} \, \sigma^z_{\zeta\eta},
\end{align}
\end{subequations}
where $\eta$ and $\zeta$ run over $\uparrow$ and $\downarrow$ and the matrices $\sigma$ are Pauli matrices.  Insisting that $X+Y+Z=U_0 + V_+ + W_-$ reveals that
\begin{subequations}
\begin{align}
u_i^a &= z_i^a,
\\
v_i^a &= x_i^a + \mathrm{i} \, y_i^a,
\\
w_i^a &= x_i^a - \mathrm{i} \, y_i^a.
\end{align}
\end{subequations}

What does this mean for the SGHF operators $C_2$, $C_3$, and $K_4$?

We begin by noting that $\kappa(A,B)$ is bilinear:
\begin{subequations}
\begin{align}
\kappa(A +  B,C + D) 
 &= \kappa(A,C) + \kappa(A,D)
\\
 &+ \kappa(B,C) + \kappa(B,D),
\nonumber
\\
\kappa(\alpha \, A, \beta \, B) &= \alpha \, \beta \, \kappa(A,B),
\end{align}
\end{subequations}
where $\alpha$ and $\beta$ are numbers and $A$, $B$, $C$, and $D$ are operators.  It is also symmetric: $\kappa(A,B) = \kappa(B,A)$.  Together, these imply
\begin{equation}
\kappa(V,W) = \kappa(X + \mathrm{i} \, Y, X - \mathrm{i} \, Y) = \kappa(X,X) + \kappa(Y,Y)
\end{equation}
and hence
\begin{equation}
C_2 = \kappa(X,X) + \kappa(Y,Y) + \kappa(Z,Z).
\end{equation}
Let us define a rank two tensor operator $\kappa$ with elements
\begin{equation}
\kappa_{ij} = \kappa(e_i,e_j)
\end{equation}
where $e_1 = X$, $e_2 = Y$, and $e_3 = Z$.  Then we have simply
\begin{equation}
C_2 = \mathrm{Tr}(\bm{\kappa})
\end{equation}
which is of course an invariant under rotations which mix $X$, $Y$, and $Z$.  Similarly, one finds
\begin{subequations}
\begin{align}
K_4
 &= \frac{3}{5} \, \Big[\kappa(X,X) \, \kappa(Y,Y) - \kappa(X,Y)^2
\\
 &\qquad + \kappa(Y,Y) \, \kappa(Z,Z) - \kappa(Y,Z)^2
\nonumber
\\
 &\qquad + \kappa(Z,Z) \, \kappa(X,X) - \kappa(Z,X)^2\Big]
\nonumber
\\
 &= \frac{3}{10} \, \left[\mathrm{Tr}(\bm{\kappa})^2 - \mathrm{Tr}(\bm{\kappa}^2)\right],
\end{align}
\end{subequations}
which is also an invariant under these rotations.

Writing $C_3$ in terms of $X$, $Y$, and $Z$ gives us
\begin{align}
C_3 = \frac{\mathrm{i}}{3} \, \Big[&X_0 \, \left(Y_- \, Z_+ - Y_+ \, Z_-\right)
\\
 +& Y_0 \, \left(Z_- \, X_+ - Z_+ \, X_-\right)
\nonumber
\\
 +& Z_0 \, \left(X_- \, Y_+ - X_+ \, Y_-\right)\Big],
\nonumber
\end{align}
where we have defined $X_0$, $X_+$, $X_-$, and so on in analogy with our other operators.  Writing a rank three tensor with elements
\begin{equation}
\kappa_{ijk} = (e_i)_0 \, (e_j)_- \, (e_k)_+,
\end{equation}
we see that
\begin{equation}
C_3 = \frac{\mathrm{i}}{3} \, \epsilon_{ijk} \, \kappa_{ijk}
\end{equation}
which is again rotationally invariant.

\begin{figure*}[t]
\includegraphics[width=0.48\textwidth]{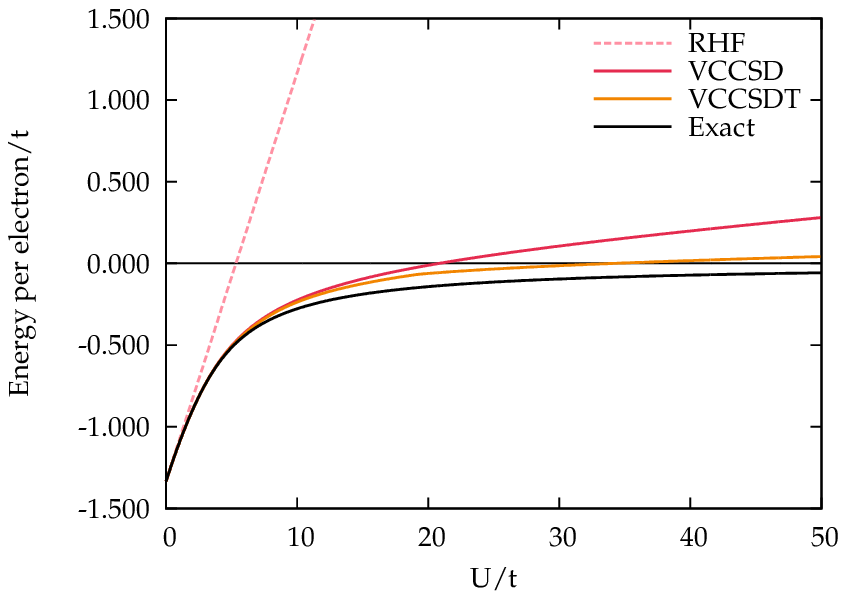}
\hfill
\includegraphics[width=0.48\textwidth]{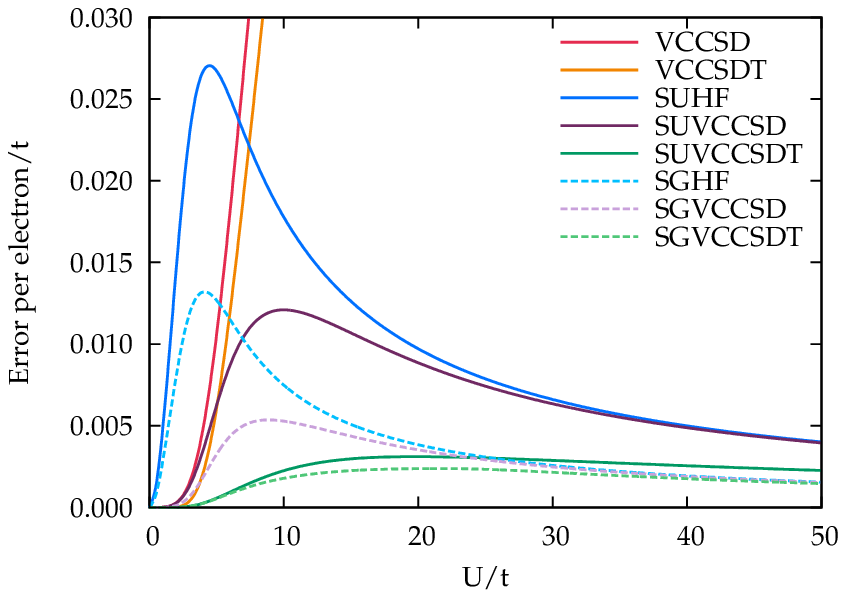}
\caption{Energies in the periodic one-dimensional 6-site Hubbard lattice at half-filling.  Left panel: Total energies per electron.  We have excluded the spin-projected methods as they are all very close to the exact result.  Right panel: Errors per electron with respect to the exact result.  Note the enormous improvement in going from SUHF and its SUVCC derivatives to SGHF and its SGVCC derivatives.
\label{Fig:Hubbard6}}
\end{figure*}

Our real point here is that the SGHF wave function can be expressed in a manifestly symmetric form in terms of simple invariants of a rank two Cartesian singlet tensor (with elements $\kappa_{ij}$) and a rank three Cartesian singlet tensor (with elements $\kappa_{ijk}$).

\section{Combining SGHF and Variational Coupled Cluster}
Our main goal in this work is to present the particle-hole form of the SGHF wave function.  Since, however, the reason we are interested in this form is so that we can combine SGHF and coupled cluster theory, we would be remiss if we did not try at least a preliminary investigation of this combination.  As in Ref. \onlinecite{Qiu2017} we use a full configuration interaction code to test the combination of SGHF with variational coupled cluster with singles and doubles\cite{Noga1988,VanVoorhis2000,Knowles2010,Evangelista2011} (VCCSD) and with variational coupled cluster with singles, doubles, and triples (VCCSDT).  In analogy with Ref. \onlinecite{Qiu2017}, we will denote the combination of SGHF and VCCSD as SGVCCSD, and the combination of SGHF with VCCSDT as SGVCCSDT; similarly, the combination of SUHF with VCCSD or with VCCSDT will be denoted respectively by SUVCCSD and SUVCCSDT.

We should emphasize that it is not our intention to use methods like SGVCCSD for production-level calculations.  Rather, we just want to see to what extent the method has promise; practical calculations would have to combine SUHF or SGHF with a more traditional similarity-transformed coupled cluster approach, which would lead to polynomial computational scaling rather than the combinatorial scaling forced by the use of variational coupled cluster.

We carry out these tests on the periodic Hubbard Hamiltonian in one dimension.  The Hamiltonian models electrons on a lattice, and is given by
\begin{equation}
H = -t \sum_{\langle ij \rangle, \sigma} \left(a_{i_\sigma}^\dagger \, a_{j_\sigma} + h.c.\right) + U \, \sum_i a_{i_\uparrow}^\dagger \, a_{i_\downarrow}^\dagger \, a_{i_\downarrow} \, a_{i_\uparrow}.
\end{equation}
Here, $i$ and $j$ index lattice sites and the notation $\langle ij \rangle$ means that sites $i$ and $j$ are connected in the lattice.  We consider only nearest-neighbor hopping ($i = j \pm 1$) and periodic boundary conditions (sites $1$ and $N$ in an $N$-site lattice are also connected).  The parameter $U/t$ controls the strength of correlation, with small $U/t$ corresponding to a weakly-correlated system and large $U/t$ corresponding to strong correlation.

We begin with the half-filled 6-site lattice, shown in Fig. \ref{Fig:Hubbard6}.  The left panel shows the total energy per electron for RHF, VCCSD, and VCCSDT in comparison to the exact result; we omit results from spin-projected methods as they are all essentially superimposable with the exact data.  To see how well the spin-projected methods perform, particularly in combination with VCCSD or VCCSDT, we show the error per electron with respect to the exact result in the right panel.  It is readily apparent that in this half-filled case, neither VCCSD nor even VCCSDT is of particularly high quality except at small $U/t$.  For sufficiently large $U/t$, SUHF is very good.  Adding VCCSD to SUHF improves the results significantly in the intermediate range of $U/t$ but has little effect for large $U/t$; this deficiency is remedied by adding triple excitations to yield SUVCCSDT.  Remarkably, SGHF is already superior to SUVCCSD for most $U/t$, and SGVCCSD is better even than SUVCCSDT for the strongly correlated limit.  Unlike when we add coupled-cluster correlations to SUHF, triple excitations appear to add no meaningful corrections to SGVCCSD for large $U/t$ though they are naturally important in the less strongly correlated limit.  This suggests that the dominant effect of going from SUVCCSD to SUVCCSDT is that the triple excitations we add mimic the effect of the spin-projected GHF (which, of course, contains a triply-excited component).  Note that lattice momentum symmetry eliminates single excitations in this problem, so SUVCCSD contains only even excitation levels.  We should also point out that in the $U/t \to \infty$ limit, UHF becomes energetically exact; consequently so too do SUHF, SGHF, and their combinations with variational coupled cluster theory.

\begin{figure*}[t]
\includegraphics[width=0.48\textwidth]{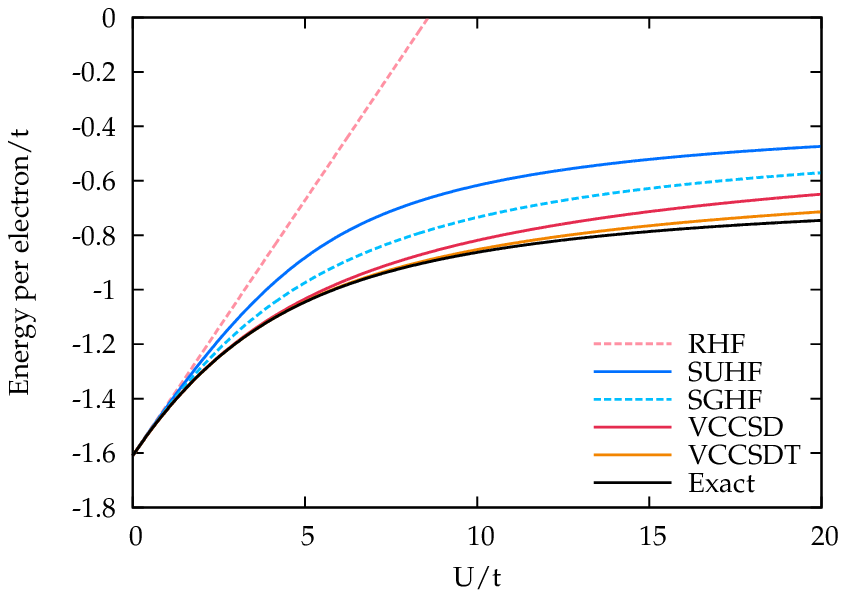}
\hfill
\includegraphics[width=0.48\textwidth]{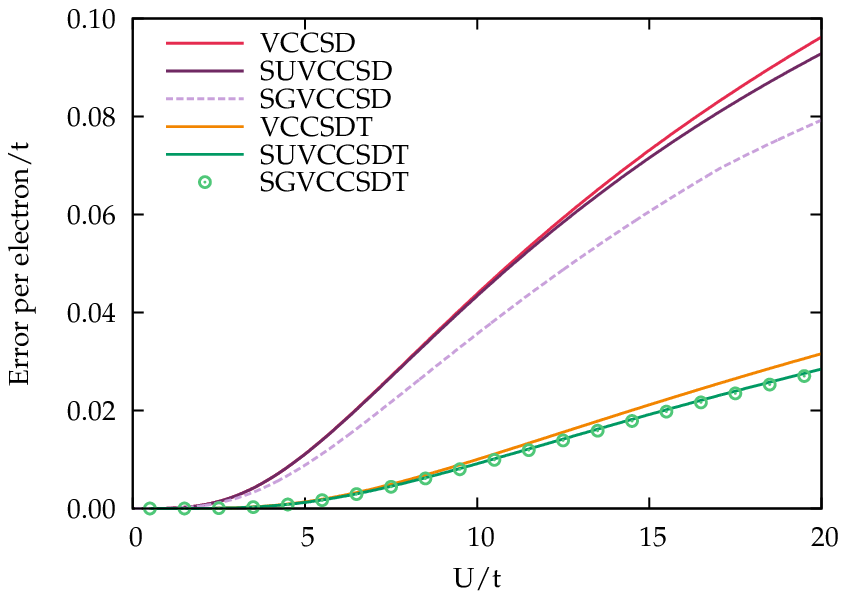}
\caption{Energies in the periodic one-dimensional 8-site Hubbard lattice with six electrons.  Left panel: Total energies per electron.  We have not shown SUVCC or SGVCC, which are virtually indistinguishable from the variational coupled cluster results.  Right panel: Errors per electron with respect to the exact result.  We have excluded SUHF and SGHF, which are clearly much poorer than VCCSD for this system.
\label{Fig:Hubbard8}}
\end{figure*}

While adding SUHF or particularly SGHF to variational coupled cluster provides significant improvements over variational coupled cluster theory alone in the half-filled lattice, this is not always true.  The half-filled Hubbard lattice is something of a best case scenario for methods like SGVCC, as variational coupled cluster breaks down here while SUHF or SGHF are exceptionally accurate in the strongly correlated limit.  On the other hand, consider the 8-site lattice with six electrons, as shown in Fig. \ref{Fig:Hubbard8}.  Here, SUHF and SGHF, while significantly improving upon RHF, are no longer energetically exact in the large $U/t$ limit (i.e. they do not capture all strong correlations).  In contrast, variational coupled cluster theory is already quite good, and is in fact better than SUHF or SGHF for all $U/t$.  The result is that while adding SGHF to VCCSD provides a modest amount of improvement, essentially nothing is gained by adding SUHF, and VCCSDT is so close to the exact result that even adding SGHF to it has minimal effect.  We should also point out that generally, our errors with respect to the exact result are significantly larger in this doped system.

\section{Symmetry Projection of More Elaborate Wave Functions}
Traditionally, efficient formulations of symmetry projection have started by integrating the broken-symmetry wave function over a manifold of coherent states.\cite{Ring80,Blaizot85,Schmid2004,PHF}  While conceptually and computationally simple, this approach does not readily lend itself to combination with symmetry-preserving particle-hole excitation operators simply because the traditional formalism is based on combining broken-symmetry states.  For example, consider adding coupled cluster to spin-projected UHF.  If the cluster operator $T_2$ were expressed in terms of broken-symmetry orbitals, it would itself break symmetry, so one would need to spin project the unrestricted coupled cluster wave function.\cite{Duguet2014,Duguet2016,Qiu2017b}  If, on the other hand, the cluster operator were expressed as symmetry-preserving excitations, it would not have a particle-hole form in the basis of the broken-symmetry orbitals.  

It is therefore important to ascertain how one could write the symmetry-projected wave function in terms of the symmetry-adapted orbitals.  Unfortunately, for each symmetry (and possibly for each symmetry eigenvalue!) the particle-hole representation of the symmetry-projected state is different.  This certainly presents a practical challenge in actually working out the appropriate projected state.  While to date we have found that symmetry-projected wave functions can be written as polynomials of a small number of elementary excitation operators, we do not guarantee that this is always the case; if it is not, symmetry projection is further complicated by the need to identify increasingly cumbersome operators to account for increasingly higher excitation levels.

Nevertheless, the general program seems clear: one should find a Thouless transformation which transforms the symmetry-adapted reference determinant into the broken-symmetry determinant, and by working in the symmetry-adapted basis, one can with sufficient exertion carry out the symmetry projection analytically.  We have derived but not implemented the first few terms in the number- and spin-projected generalized HFB wave function.  Simpler is the form for a combination of GHF spin symmetry breaking and singlet-paired number symmetry breaking.  That is, if the Thouless transformation creating the broken-symmetry mean-field is
\begin{equation}
|\Phi\rangle = \mathrm{e}^{T_1 + U_0 + V_+ + W_- + \Sigma_{+} + \Sigma_{-}} |\mathrm{RHF}\rangle
\label{Eqn:ManyThouless}
\end{equation}
where
\begin{subequations}
\begin{align}
\Sigma_{+} &= \sum_{ab} \sigma_{ab} \, \left(c_{a_\uparrow}^\dagger \, c_{b_\downarrow}^\dagger -  c_{a_\downarrow}^\dagger \, c_{b_\uparrow}^\dagger\right),
\\
\Sigma_{-} &= \sum_{ij} \sigma_{ij} \, \left(c_{j_\uparrow} \, c_{i_\downarrow} - c_{j_\downarrow} \, c_{i_\uparrow}\right),
\end{align}
\end{subequations}
then the number- and spin-projected wave function is just
\begin{equation}
P |\Phi\rangle = \sum \frac{1}{(n!)^2} \, \lambda_{ijk} \, \left(\Sigma_{+} \, \Sigma_{-}\right)^n \, C_2^i \, C_3^j \, K_4^k |\mathrm{RHF}\rangle.
\end{equation}
That is, because $\Sigma_{+}$ and $\Sigma_{-}$ preserve spin symmetry while $U_0$, $V_+$, and $W_-$ preserve number symmety, the wave function is just the product of the SGHF spin projection polynomial and the number-projected HFB polynomial.  In obtaining this result we have made use of the fact that the various Thouless transformation operators of Eqn. \ref{Eqn:ManyThouless} are all mutually commuting.

Perhaps more interesting yet, one can write a more general broken-symmetry coupled-cluster-style wave function in terms of symmetry breaking cluster operators acting on a symmetry-adapted determinant, and then project the result.  For example, consider a number-projected case for simplicity, in which we write
\begin{equation}
|\Psi\rangle = P_N \, \mathrm{e}^{T_{+1} + T_{-1} + T_{+2} + T_{-2}} |\mathrm{RHF}\rangle
\end{equation}
where $T_{\pm k}$ adds or removes $2k$ electrons from the system.  The wave function would start as
\begin{align}
|\Psi\rangle &= \left(1 + T_{+1} \, T_{-1} + \frac{1}{4} \, T_{+1}^2 \, T_{-1}^2 + T_{+2} \, T_{-2}\right.
\\
 &\qquad + \left.\frac{1}{2} \, T_{+2} \, T_{-1}^2 + \frac{1}{2} \, T_{-2} \, T_{+1}^2 + \ldots \right) |\mathrm{RHF}\rangle
\nonumber
\end{align}
and would clearly be more complete than the simple number-projected HFB in which $T_{\pm 2}$ are neglected.

\section{Conclusions}
Incorporating symmetry projection and methods to describe dynamic correlation has been a long-standing goal of many researchers.  This combination is greatly simplified by writing the symmetry-projected mean-field determinant in terms of particle-hole excitations out of some symmetry-adapted reference determinant.  The particle-hole representation of number projected BCS is textbook material (see, e.g., Eqn. 8.196 of Ref. \onlinecite{Ring80}).  The particle-hole form of singlet spin-projected UHF has also been recognized for quite some time,\cite{Piecuch1996} though only recently has it been realized that the SUHF wave function can be written as a polynomial of single and double excitation operators.\cite{Qiu2016,Qiu2017}  This work shows that the singlet SGHF also has such a polynomial form, though now we need not only single excitation operators ($T_1$) and double excitation operators ($C_2$) to define the polynomial, we must also include triple $(C_3)$ and quadruple excitation operators ($K_4$), but apparently no further.  These higher excitations appear only once the Thouless transformation from RHF to symmetry-broken Hartree-Fock has more than one mode (i.e. we require $U_0$, $V_+$, and $W_-$).

Having a representation of a projected Hartree-Fock wave function in terms of symmetry-adapted wave functions opens the way to several possible combinations of projected Hartree-Fock and coupled cluster theory.  Here and in Ref. \onlinecite{Qiu2017} we have explored the simple wave function ansatz $|\Psi\rangle = \exp(T) |\mathrm{PHF}\rangle$ in a variational framework as a precursor to more computationally useful approaches which use the correlator $\exp(T)$ to define a similarity-transformed Hamiltonian as in traditional coupled cluster theory.  Our work on attenuated coupled cluster\cite{Gomez2017} follows similar ideas.  Much earlier work explored writing the PHF wave function as an exponential of higher-order cluster operators (e.g. $|\mathrm{SUHF}\rangle = \exp(T_2 + T_4 + T_6 + \ldots) |\mathrm{RHF}\rangle$) and taking advantage of this structure, perhaps in combination with the factorized form of the operators $T_n$ obtained from PHF wave functions, to obtain accurate results for strongly correlated systems at the cost of a well-defined wave function.\cite{Piecuch1990,Piecuch1991,Paldus1992,Piecuch1996}  Most of this work to date has focused on merging SUHF and coupled cluster in some way, but it could be generalized to the SGHF case.

Clearly there is quite a lot of work to be done before we can optimally combine projected mean-field states and coupled cluster theory in practical calculations.  Early returns, however, seem promising, and we have high hopes that in the fullness of time these methods will permit accurate and essentially black-box calculations for both strongly and weakly correlated systems.

\begin{acknowledgments}
This work was supported by the U.S. Department of Energy, Office of Basic Energy Sciences, Computational and Theoretical Chemistry Program under Award No. DE-FG02-09ER16053. G.E.S. is a Welch Foundation Chair (C-0036).  TMH would like to thank Yiheng Qiu for helpful discussions and John Gomez and Carlos Jim\'enez-Hoyos for providing benchmark SGHF data for us to compare to.
\end{acknowledgments}

\bibliography{SGHF}
\end{document}